# Three-body Förster resonance of a new type in Rydberg atoms

P.Cheinet [1], K.-L.Pham [1], P.Pillet [1], I.I.Beterov [2,3,4], I.N.Ashkarin [2,3], D.B.Tretyakov [2,3], E.A.Yakshina [2,3], V.M.Entin [2,3], I.I.Ryabtsev [2,3]

[1] *Laboratoire Aime Cotton, CNRS, Univ. Paris-Sud, ENS-Cachan, Universite Paris-Saclay, 91405 Orsay, France*

[2] *Rzhanov Institute of Semiconductor Physics SB RAS Prospekt Lavrentyeva 13, 630090, Novosibirsk, Russia*

[3] *Novosibirsk State University Ul. Pirogova 2, 630090, Novosibirsk, Russia*

[4] *Novosibirsk State Technical University Prospekt Marksa 20, 630073, Novosibirsk, Russia*

***Abstract.*** The three-body Förster resonances $3 \times nP_{3/2} \to nS_{1/2} + (n+1)S_{1/2} + nP_{3/2}^*$ controlled by a dc electric field were realized earlier by the authors in an ensemble of several cold Rydberg Rb atoms. One of the drawbacks of such resonances for potential application in three-qubit quantum gates is the proximity of the two-body Förster resonance $2 \times nP_{3/2} \to nS_{1/2} + (n+1)S_{1/2}$, as well as the possibility of their implementation only for states with values of the principal quantum numbers $n \leq 38$. In this paper we propose and analyze a three-body resonance of a new type $3 \times nP_{3/2} \to nS_{1/2} + (n+1)S_{1/2} + nP_{1/2}$, which can be realized for arbitrary $n$. Its specific feature is also that the third atom transits into a state with a different total angular moment $J=1/2$, which has no Stark structure, so that the two-body resonance is completely absent. Numerical calculations showed that for not too strong interaction, it is possible to observe coherent three-body oscillation of the populations of collective states, which is of interest for developing new schemes of three-qubit quantum gates controlled by an electric field.

## 1. Introduction

Atoms in highly-excited Rydberg states with principal quantum numbers $n \gg 1$ have large dipole moments growing like $n^2$ and can experience strong long-range interactions [1]. This is especially attractive for developing quantum computers and simulators with qubits based on single alkali-metal atoms trapped in arrays of optical dipole traps or in optical lattices [2, 3]. In particular, quantum simulators based on Rydberg atoms can directly simulate various objects in solid-state physics due to the ability to include all possible interactions between their components, if such interactions in a quantum simulator are appropriately controlled [4].

The interactions between Rydberg atoms can be controlled using a dc electric field. Since the polarizabilities of Rydberg states grow as $n^7$, even a weak electric field causes large Stark shifts of the energies of Rydberg states. As an example, Fig. 1a shows the calculated Stark diagram of Rydberg states of Rb atoms near the 37$P$ state. Rydberg $S$, $P$, and $D$ states have large quantum defects and experience the quadratic Stark effect. States with large orbital angular momenta ($L > 2$) are degenerate in energy and experience the linear Stark effect.

Electrically tuned Förster resonances correspond to Förster resonance energy transfer [5]. They arise due to the crossing of collective energy levels of Rydberg atoms at a certain value of the electric field. Förster resonances are controlled by the electric field over the interaction strength and distance and can be either a resonant dipole-dipole interaction (with exact resonance) or non-resonant van der Waals interaction (with a large detuning from the resonance) [6]. Förster resonances can be either two-body, when the states of only two atoms in the ensemble change due to interactions, or many-body, when the states of more than two atoms change simultaneously. This is of interest, e.g., for the implementation of three-qubit quantum gates used in error correction algorithms in quantum computing [7].

As an example, Fig. 1b shows the calculated Stark structure of the Förster resonance $3 \times 37P_{3/2}(|M|) \rightarrow 37S_{1/2} + 38S_{1/2} + 37P_{3/2}(|M^*|)$ for three Rydberg Rb atoms ($|M|$ is the projection of the angular moment $J$ on the quantization axis $Z$). The dependences of the energies $W$ of various three-body collective states on the control electric field are presented. The crossings of collective states 2–7 are actually two-body resonances that do not require the presence of a third atom. Crossings 1 and 8 correspond to three-body resonances, possible only in the presence of a third atom, which takes away an energy defect that impedes two-body resonance. Figure 1c shows a simplified diagram of the three-body Förster resonance for three Rydberg atoms. The initial collective state is state 1. Intermediate state 2 corresponds to two atoms in the $S$-states and one atom in the initial $P$-state. The final state is state 3, which has a different projection of the atomic angular moment in the $P$-state. The detunings $\Delta_1$ and $\Delta_2$ are controlled by the electric field. Three-body resonance occurs at $\Delta_1 = \Delta_2$, while two-body resonance at $\Delta_1 = 0$.

Three-body Förster resonances were first proposed and realized in a gas of cold Rydberg Cs atoms [8]. In such three-body resonances, one of the atoms takes away an excess of energy, which prevents the two-body process; this leads to the Förster energy transfer of the so called 'Borromean' type. A Borromean transition is characterized by a strong isolated three-body energy transfer with a negligible contribution from the two-body effect. This allows studying the three-body effect, which is usually impossible to observe in other systems, since it turns out to be hidden by the strong signal from the two-body effect. The experiment [8] was performed with an ensemble of ~$10^5$ Cs atoms in the interaction volume with a characteristic size of ~200 µm. Therefore, a three-body resonance was actually observed for a large number of atoms $i \gg 1$.

We recently performed a similar experiment for $i$ = 1 - 5 Rydberg Rb atoms randomly located in the volume of laser excitation with a characteristic size of ~15 µm [9]. Figure 1d shows experimental records of the Förster resonances $\rho_i$ observed for different numbers of atoms and their initial states $37P_{3/2}(|M|=1/2)$ or $37P_{3/2}(|M|=3/2)$ at the interaction time 3 µs ($\rho_i$ is the population of the final collective state for a certain number of interacting atoms $i$). Two-body resonances 3 and 6 are absent for $i$ = 1, which confirms their two-body nature. Three-body resonances 1 and 8 are absent for $i$ = 1, 2, which confirms their three-body nature. Thus, clear evidence was found that the three-body resonance $3 \times 37P_{3/2}(|M|) \rightarrow 37S_{1/2} + 38S_{1/2} + 37P_{3/2}(|M^*|)$ for $n$ = 36, 37 does not manifest itself for two interacting Rydberg atoms, while it is present for three or more atoms. A theoretical analysis of three-body Förster resonances was performed by us in Ref. [10], where it was shown that for fixed positions of three atoms it is possible to observe coherent oscillations in the populations of interacting atoms and control the phase of the collective wave function. Based on such three-body resonances, we proposed a scheme for executing the three-qubit quantum Toffoli gate with a fidelity exceeding 98 % [11].

The above three-body Förster resonances $3 \times 37P_{3/2}(|M|) \rightarrow 37S_{1/2} + 38S_{1/2} + 37P_{3/2}(|M^*|)$ corresponded to the two-stage process $3 \times nP_{3/2}(|M|) \rightarrow nS_{1/2} + (n+1)S_{1/2} + nP_{3/2}(|M|) \rightarrow nS_{1/2} + (n+1)S_{1/2} + nP_{3/2}(|M^*|)$, in which each of the stages was not energy resonant, and there was resonance only between the initial and final collective states (Fig. 1c). For such resonances, two of the three interacting atoms in the initial state $nP_{3/2}(|M|)$ transited to the neighboring states $nS_{1/2}$ and $(n+1)S_{1/2}$, and the third atom remained in the state $nP_{3/2}(|M|)$, but changed the projection of the angular moment (if the initial projection was $|M|$=1/2, then it changed to $|M^*|$=3/2, and vice versa). The usual two-body Förster resonance corresponds to the case $2 \times nP_{3/2} \rightarrow nS_{1/2} + (n+1)S_{1/2}$, when the third atom does not change the projection of the angular moment; therefore, this resonance occurs in a slightly different electric field.

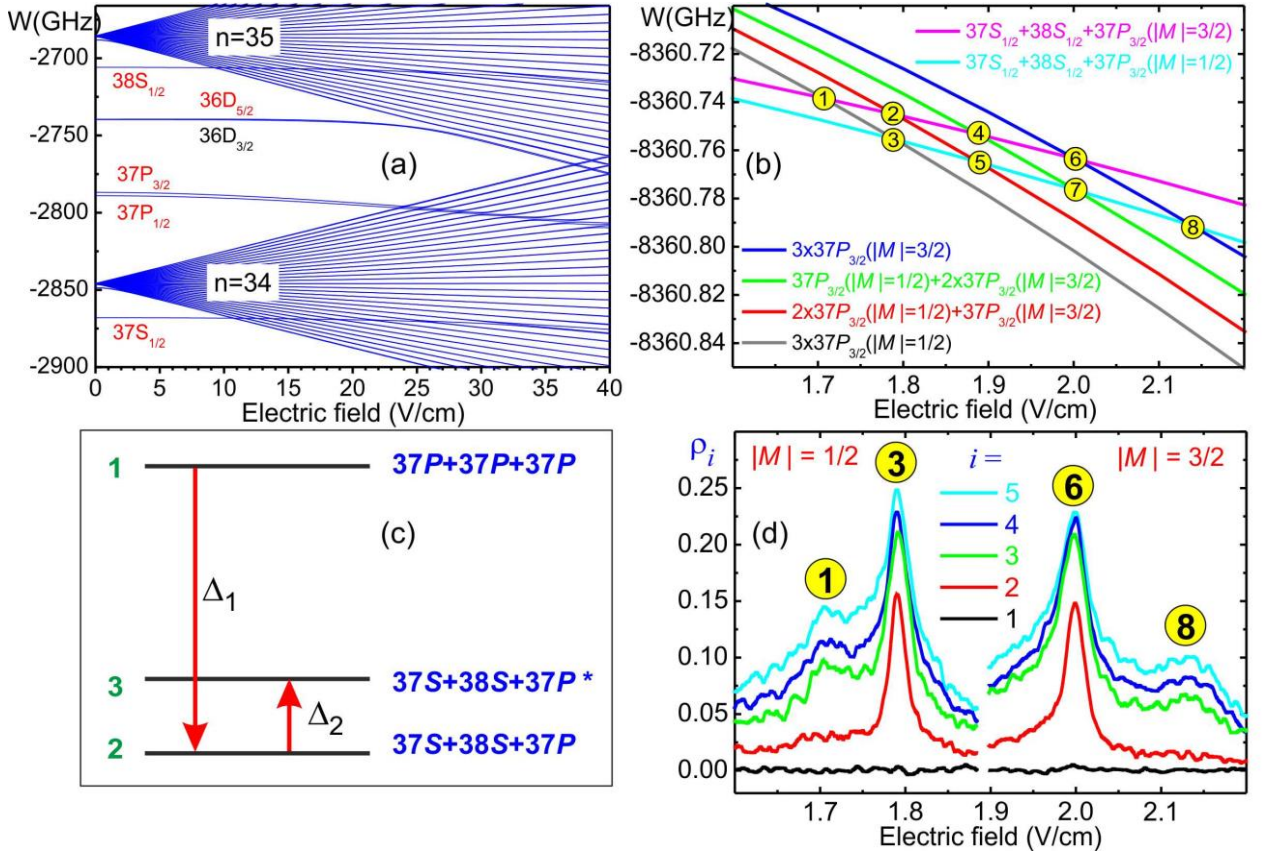

**Figure 1.** (a) Calculated Stark diagram of Rydberg states of Rb atoms with the projection of the angular moment $|M|=1/2$ near the state $37P$, (b) calculated Stark structure of the Förster resonance $3\times 37P_{3/2}(|M|) \to 37S_{1/2}+38S_{1/2}+37P_{3/2}(|M^*|)$ for three Rydberg Rb atoms, (c) simplified scheme of the three-body Förster resonance for three Rydberg atoms, and (d) experimentally recorded Förster resonances $\rho_i$ observed for different numbers of atoms ($i = 1 – 5$) and their initial states $37P_{3/2}(|M|=1/2)$ or $37P_{3/2}(|M|=3/2)$ at an interaction time of 3 μs in a single excitation volume with a characteristic size of ~15 μm and random arrangement of atoms. Numbers $1 – 8$ denote the crossings of collective states corresponding to Förster resonances of various natures.

The first transition is an ordinary two-body resonance, while the second transition occurs due to a non-resonant exchange interaction corresponding to excitation jumps between Rydberg atoms in the *S*- and *P*-states. The first and second transitions occur simultaneously, which implies their Borromean nature, while the third atom compensates for the energy of the nonzero defect of the two-body Förster resonance. Therefore, three-body resonances are less efficient than two-body resonances in the case of a weak dipole-dipole interaction. However, when the three-body resonance is precisely tuned by the electric field, its contribution to the population transfer generally exceeds the contribution of the two-body interaction, which in this case turns out to be non-resonant. Thus, the condition of the Borromean nature of three-body interactions is fulfilled [10].

In the case of Rb Rydberg atoms, one of the drawbacks of the above three-body resonances $3\times 37P_{3/2}(|M|) \to 37S_{1/2}+38S_{1/2}+37P_{3/2}(|M^*|)$ when used to implement three-qubit quantum gates is the proximity of the two-body resonance $2\times nP_{3/2} \to nS_{1/2}+(n+1)S_{1/2}$, which partially overlaps with the three-body one [9]. Another drawback is that, due to the specific values of quantum defects and polarizabilities of Rydberg states *nP* and *nS* in Rb atoms, the crossing of collective energy levels in a control electric field corresponding to three-body Förster resonances is possible only for states with values of the principal quantum number $n \leq 38$. However, to

increase the accuracy of quantum operations, it is necessary to perform them with higher Rydberg states with long radiative lifetimes and large transition dipole moments.

Therefore, the aim of this work was to search for other possible three-body resonances for Rb Rydberg atoms in the states *nP* and *nS*, which have a simple structure of Stark sublevels. States with higher orbital angular moments have a complex Stark structure with many sublevels (Fig. 1a) and generally are not suitable for high-fidelity quantum gates.

## 2. Three-body Förster resonances in an ensemble of three Rydberg atoms

A theoretical analysis of three-body Förster resonances $3 \times 37P_{3/2}(|M|) \rightarrow 37S_{1/2} + 38S_{1/2} + 37P_{3/2}(|M^*|)$ was performed by us in Ref. [10]. For the transition diagram in Fig. 1, in the case of three fixed Rydberg atoms in a triangular configuration, the following analytical solution was obtained for the line shape and time evolution of three-body Förster resonances:

$$\rho_3 \approx \frac{\Omega_0^2/3}{(\Delta - \Delta_0)^2 + \Omega_0^2} \sin^2\left[\frac{t}{2}\sqrt{(\Delta - \Delta_0)^2 + \Omega_0^2}\right], \quad (1)$$

where $\Delta = \Delta_1 - \Delta_2$ is the detuning from the unperturbed three-body Förster resonance; $\Delta_0 = -2\Omega_2 + (4\Omega_2^2 - 6\Omega_1^2)/(\Delta_2 + 2\Omega_1)$ is the dynamic shift of the three-body resonance due to non-resonant intermediate interactions with the matrix elements of the dipole-dipole interaction operators $\Omega_1 = V_1/\hbar$ and $\Omega_2 = V_2/\hbar$ for the transitions $1 \rightarrow 2$ and $2 \rightarrow 3$, correspondingly; $V_1$ and $V_2$ are the dipole-dipole interaction energies; and $\Delta_0 = -2\Omega_2 + (4\Omega_2^2 - 6\Omega_1^2)/(\Delta_2 + 2\Omega_1)$ is the oscillation frequency of the populations of collective states at exact resonance. This formula is identical to Rabi oscillations for a two-photon transition in a three-level system with a detuned intermediate level 2, which is not populated, and population oscillations occur only between levels 1 and 3. The required interaction time *t* for long-lived Rydberg states can be set either by the Stark switching method in a pulsed electric field, as in our work [9], or by the time between laser excitation and the subsequent de-excitation of given Rydberg states. From Eq. (1), several important conclusions can be drawn.

First, the three-body resonance experiences a dynamic shift $\Delta_0$, consisting of two parts: the part with $-2\Omega_2$ arises due to always-resonant exchange interactions of atoms in the *S*- and *P*-states, and the other part is a dynamic Stark shift caused by intermediate non-resonant interactions. Therefore, the position of the three-body resonance in the scale of the controlling electric field depends on the interaction energy and the ratio of the energies of the intermediate transitions $1 \rightarrow 2$ and $2 \rightarrow 3$. In real Rydberg atoms, there is a Zeeman structure of Rydberg levels, which gives rise to multiple interaction channels with various matrix elements of dipole-dipole interaction. Therefore, due to the dynamic shift, they should demonstrate a set of individual resonances arising at slightly different values of the electric field rather than single three-body resonances. If the difference is large enough, it is possible to manipulate with individual channels of interaction. By choosing the spatial configuration of the three atoms, some channels can be suppressed. Note that the line shape defined by Eq. (1) for a fixed interaction time differs from what we observed in the experiment (Fig. 1d) in a single excitation volume with a characteristic size of ~15 μm and random arrangement of atoms. This is because for a random arrangement of atoms all interaction channels are realized. This leads to an effective broadening of the three-body resonance.

Second, for immobile atoms, Eq. (1) demonstrates the possibility of coherent oscillations of populations appearing at exact resonance ($\Delta = \Delta_0$). The frequency of these oscillations $\Omega_0$ depends on the specific interaction channel. The maximum resonance amplitude for $\rho_3$ is 1/3 (in the experiment, the amplitude is defined as the fraction of atoms in the final state $nS$, which is recorded by the selective field ionization of Rydberg atoms [9]). The resonance width is determined by the combination of the Fourier width of the interaction pulse and the value of $\Omega_0$.

Third, each minimum of population oscillations corresponds to a phase shift $\pi$ for the collective wave function. Since such oscillations are controllable and reversible, they can be used to implement three-qubit quantum gates with Rydberg atoms, in particular, the Toffoli gate [11].

At the same time, as already mentioned, the three-body Förster resonances $3 \times 37P_{3/2}(|M|) \to 37S_{1/2} + 38S_{1/2} + 37P_{3/2}(|M^*|)$ have the following disadvantage from the point of view of their usage for quantum gates: they work only for low Rydberg states and can partially overlap with two-body Förster resonances, since there are two possible final states of the triatomic system in the Stark diagram in Fig. 1b, corresponding to two- and three-body resonances.

We performed an extended analysis of other possible three-body Förster resonances tuned by an electric field upon excitation of higher Rydberg states $nP_{3/2}$. One of the possible options was the case when three atoms are excited into different initial Rydberg states rather than the same ones. For example, in Ref. [11], to implement the fast three-qubit Toffoli quantum gate we proposed to use the three-body Förster resonances for the initial collective state $|80P_{3/2}(M=+3/2); 81P_{3/2}(M=+3/2); 81P_{3/2}(M=-3/2)\rangle$ and showed the possibility of achieving high fidelity of the operation. However, this scheme is hard to implement experimentally, because it requires several different-frequency exciting laser radiations rather than one.

Therefore, in this work, we found a new simpler three-body resonance $3 \times nP_{3/2} \to nS_{1/2} + (n+1)S_{1/2} + nP_{1/2}$, in which three atoms are excited into identical Rydberg states, and there is only one final state. This resonance turned out to be realizable for arbitrary initial Rydberg states $nP_{3/2}$. As an example, Fig. 2 shows the calculated Stark map of Rydberg states of Rb atoms near the 70P state and the calculated Stark structure of the new-type Förster resonance $3 \times 70P_{3/2} \to 70S_{1/2} + 71S_{1/2} + 70P_{1/2}$ for three Rydberg Rb atoms. For this resonance, the crossings of collective states (indicated by numbers) correspond only to three-body Förster resonances, when all three atoms change their states, and there are no two-body resonances at all.

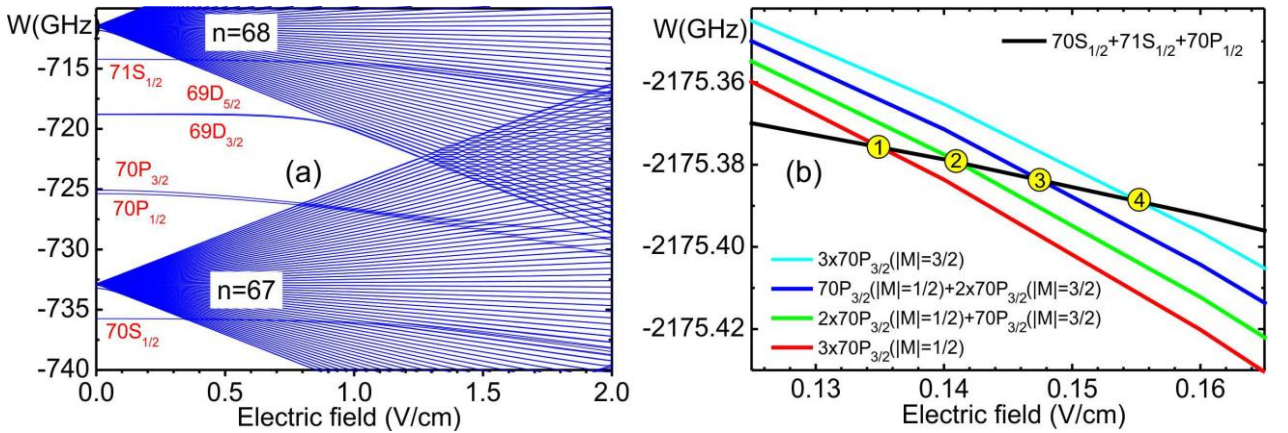

**Figure 2.** (a) Calculated Stark map of Rydberg states of Rb atoms near the 70P state and (b) calculated Stark structure of a new-type Förster resonance $3 \times 37P_{3/2} \to 37S_{1/2} + 38S_{1/2} + 37P_{1/2}$ for three Rydberg Rb atoms. Crossings of collective states (indicated by numbers) correspond only to three-body Förster resonances, when all three atoms change their states, while two-body resonances are absent.

A distinctive feature of this resonance is that the third atom transits to a state with different total angular moment $J = 1/2$ having no Stark structure rather than to a state with a different projection of the same angular moment. Therefore, the experimental study of such a three-body resonance should be much simpler, since two-body resonance is completely absent. At the same time, such a resonance may be noticeably weaker due to significantly larger detunings of intermediate levels (about 200 MHz) than in the case of the three-body resonance $3 \times 37P_{3/2}(|M|) \rightarrow 37S_{1/2} + 38S_{1/2} + 37P_{3/2}(|M^*|)$ (about 10 MHz). On the other hand, for high Rydberg states, the dipole moments of transitions are much larger. For example, for transitions from the 70P state to neighboring states 70S and 71S, the radial part of the dipole moments is about 5,000 a.u., while similar transitions from the 37P state have radial parts of about 1,300 a.u. As a result, the value of $\Omega_0$ for the resonance of a new type is of the same order of magnitude as for the resonances of the old type.

### 3. Results of numerical calculations for three-body Förster resonances of a new type

As already discussed, in real Rydberg atoms, due to the presence of several interaction channels, a set of several resonances corresponding to different channels should be observed instead of a single three-body Förster resonance. In Ref. [10], we showed that in order to reduce the number of such channels, the optimal geometry of the three atoms is their uniform location along the quantization axis Z, which is chosen along the direction of the control electric field. In this case, only two well-separated three-body Förster resonances corresponding to two interaction channels remain.

Nevertheless, analytical calculations for such a geometry turn out to be impossible; therefore, we performed numerical calculations in the same way as it was done in Ref. [10] for three-body resonances $3 \times 37P_{3/2}(|M|) \rightarrow 37S_{1/2} + 38S_{1/2} + 37P_{3/2}(|M^*|)$. However, in Ref. [10], all magnetic sublevels of Rydberg states were taken into account, which for the three-body Förster resonance in atoms in the $37P_{3/2}$ states required 160 collective states with all possible values of the angular moment projections to be taken into account. For the $70P_{3/2}$ state, such calculations would require taking into account a much larger number of collective states. Therefore, to reduce the number of basis states and save computing time, we used a simplified model in which the signs of the angular moment projections were not taken into account (i.e., a simplified model was constructed for Stark Rydberg sublevels rather than Zeeman ones). Its operability was checked earlier in calculations for the $37P_{3/2}$ state, which showed satisfactory agreement with the results of Ref. [10] regarding the positions and amplitudes of three-body resonances.

Figure 3 shows the results of numerical calculations of the three-body Förster resonance of a new type, $3 \times 70P_{3/2}(|M|=1/2) \rightarrow 70S_{1/2} + 71S_{1/2} + 70P_{1/2}$, for three Rydberg Rb atoms in several spatial configurations. Figure 3a corresponds to the case when three atoms are uniformly located along the Z-axis spaced by $R = 10$ μm, the interaction time being 0.35 μs. At such a distance, the interaction of neighboring Rydberg atoms is relatively weak, and the three-body resonances do not broaden. As expected, only two resonances arise in this configuration, which correspond to two interaction channels. Their resonance electric fields of 0.1247 and 0.140 V cm$^{-1}$ are close to the calculated value of 0.135 V cm$^{-1}$ for the crossing of collective levels in Fig. 2b in the absence of interaction, taking into account additional dynamic shifts. The resonance amplitudes tend to the maximum possible value 1/3, and their width when converted to the frequency scale corresponds to the Fourier width of the interaction pulse (about 3 MHz). The resonances are well resolved, and by tuning the electric field, one can select a specific three-body interaction channel.

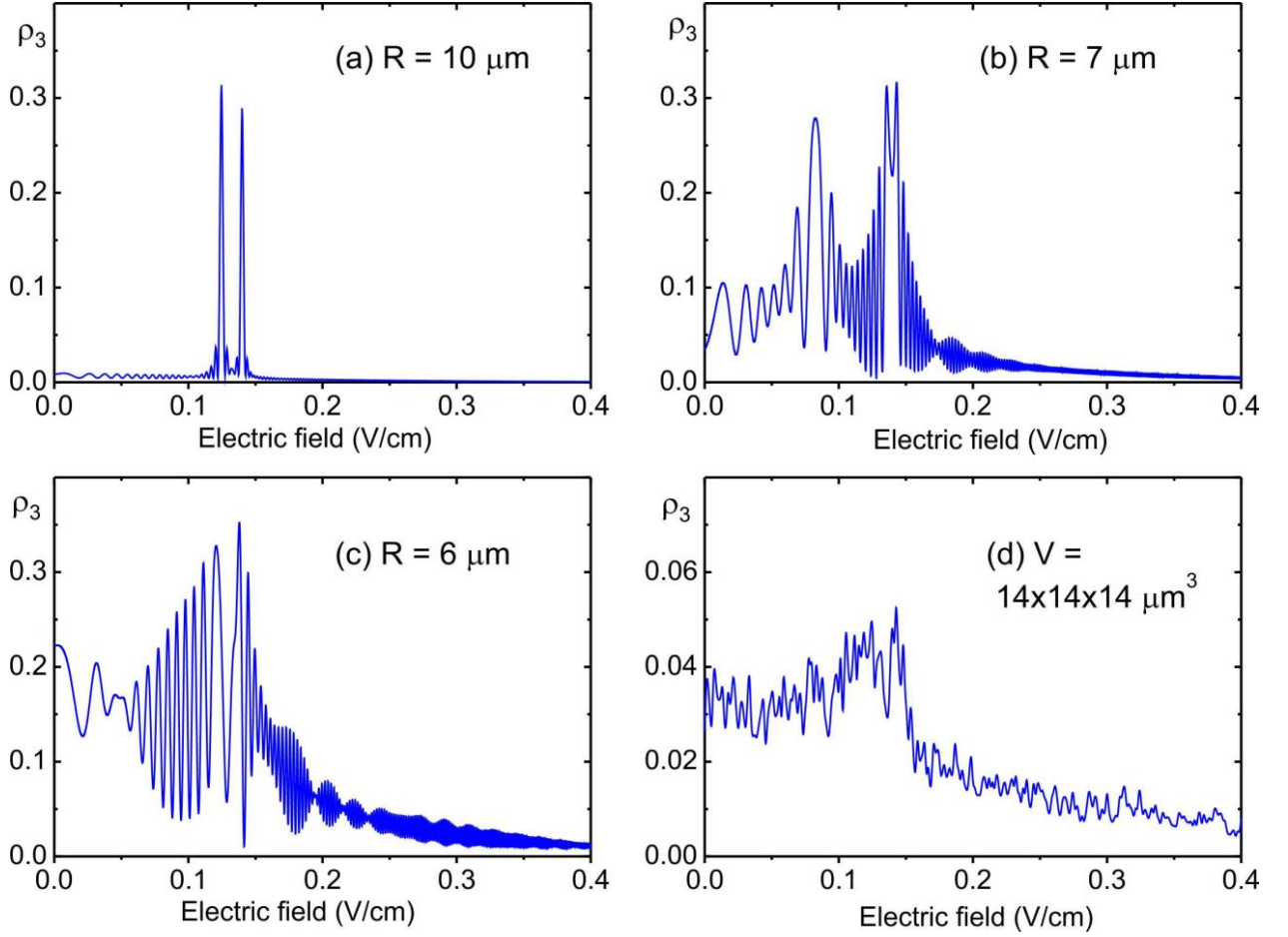

**Figure 3.** Results of numerical calculations of the three-body Förster resonance spectra of a new type $3 \times 70P_{3/2}(|M|=1/2) \to 70S_{1/2} + 71S_{1/2} + 70P_{1/2}$, for three Rydberg Rb atoms in several spatial configurations: (a) three atoms are uniformly located along the Z axis with the interatomic spacing of $R = 10$ μm at an interaction time of 0.35 μs; (b, c) the same for $R = 7$ (b) and (c) 6 μm; and (d) three atoms are randomly arranged in a cubic interaction volume $V=14\times14\times14$ μm³, averaged over 100 realizations at an interaction time of 2 μs.

With a decrease in the interatomic distance to $R = 7$ μm (Fig. 3b), the effective three-body interaction energy $\Omega_0$, which depends on the distance as $R^{-6}$, increases by 8.5 times. As a result, the calculated spectra begin to noticeably broaden, shift, and partially overlap in the presence of population oscillations. In this case, one of the resonances shifts toward a smaller electric field, and its wing has a nonzero value even in a zero field. A further decrease in the distance to $R = 6$ μm (Fig. 3c) increases the energy of three-body interaction by another 2.5 times, which is accompanied by a complete overlap of the two resonances and their considerable broadening. The observed oscillations of the populations at the wings of a three-body resonance have a period that increases with decreasing electric field, which is explained by the quadratic nature of the Stark effect in these states.

We also calculated the case when three atoms were randomly located in a cubic interaction volume $V=14\times14\times14$ μm³ averaged over 100 realizations at an interaction time of 2 μs, which approximately corresponded to the conditions of our experiment in Ref. [9] during recording three-body Förster resonances (see Fig. 1d). In this case, due to the uncertainty of the distance between the atoms and their mutual orientations, all the interaction channels are involved, the population oscillations are completely washed out, and the spectrum of the resonance approximately corresponds to the resonance envelope in Fig. 3c. The resonance amplitude also decreases noticeably, which is associated with the presence of zeros in the interaction energy for certain spatial configurations of atoms [12]. In this case, the three-body interaction is also present in the zero electric field, which can be explained by those random configurations of atoms, in which they are located close to each other and have large interaction energies comparable with the energy in Fig. 3c.

Based on the results of calculating the spectra of three-body Förster resonances of a new type (Fig. 3), we can draw the following conclusions. First, the spectra are highly sensitive to interatomic distances, and when a certain threshold value is reached, they begin to broaden, and individual interaction channels become indistinguishable. This leads to a loss of coherence and the absence of full population oscillations at high interaction energies. Second, with a random arrangement of three atoms in a single volume of laser excitation, coherence is completely lost, and the three-body interaction takes place even in a zero electric field, which complicates its observation under the conditions of our experiment [9] for old-type resonances. Therefore, experiments must be performed with single atoms in separate optical dipole traps with minimal fluctuations in their spatial position, as was done, e.g., in Ref. [13]. Third, for high Rydberg states, the values of the resonant electric field are rather small, therefore, to observe narrow three-body resonances, a high-stability source of the electric field is required, and all possible spurious fields must be carefully compensated to less than 1 mV cm$^{-1}$.

Our calculations also showed that, under the conditions of good spatial localization of atoms and minimization of the electric field noise, experimental realization of coherent population oscillations at the three-body Förster resonance is possible. Figure 4a shows a magnified image of the spectrum from Fig. 3a. There are two well-resolved peaks of three-body resonance at electric fields of 0.1247 and 0.140 V cm$^{-1}$. Fine tuning to one of the peaks (the required accuracy is ~0.1 mV cm$^{-1}$) allows switching on the coherent three-body interaction, accompanied by population oscillations. The calculated population oscillations when tuning the electric field to a three-body resonance in an electric field of 0.1247 V cm$^{-1}$ are presented in Fig. 4b. The contrast of oscillations exceeds 95 %, which allows considering them as the basis for three-qubit quantum gates, by analogy with the three-body resonances that we examined in Refs. [10, 11].

At present, we are performing more accurate theoretical calculations in the full interaction model (taking into account the Zeeman structure) in order to find an optimal Rydberg state for implementing three-qubit quantum gates based on the new type of three-body Förster resonances discussed in this paper. Note also that the many-body electrically controlled Förster resonances for large ensembles of Rydberg atoms were studied experimentally in recent works [14, 15], in which the possibility of observing four-particle and higher resonances was noted, which, however, requires significantly higher interaction energies.

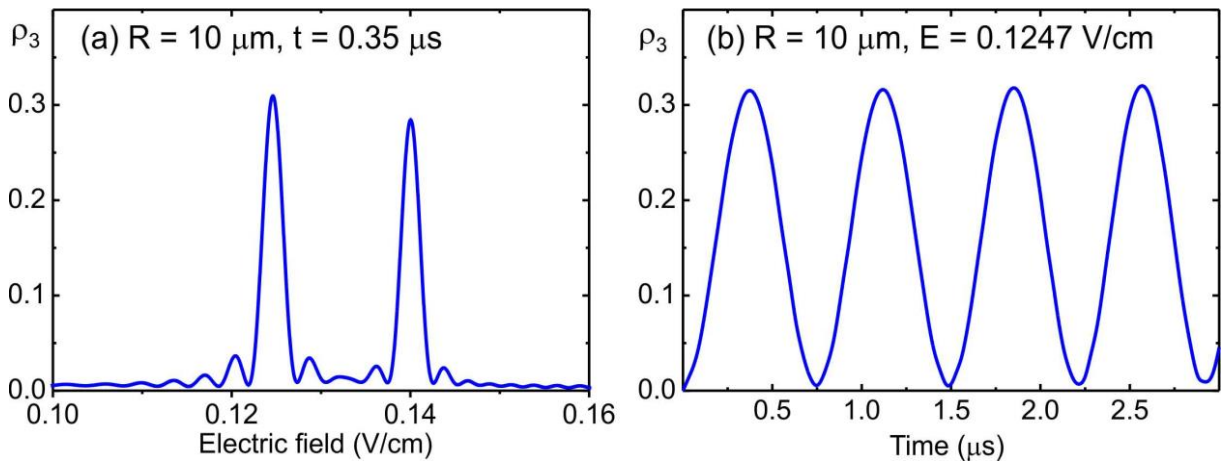

**Figure 4.** (a) Magnified image of the spectrum shown in Fig. 3a and (b) population oscillations when the electric field is tuned to a three-body resonance in an electric field of 0.1247 V cm$^{-1}$.

## 4. Conclusion

Three-body Förster resonances of a new type, $3 \times nP_{3/2} \to nS_{1/2} + (n+1)S_{1/2} + nP_{1/2}$, which can be implemented with Rb Rydberg atoms in arbitrary $nP_{3/2}$ states, have been theoretically investigated. Unlike other three-body resonances $3 \times 37P_{3/2}(|M|) \to 37S_{1/2} + 38S_{1/2} + 37P_{3/2}(|M^*|)$ studied by us previously and observed only for low-lying states with $n \leq 38$, such resonances can be observed for arbitrary states. One more specific feature of these resonances is that the third atom transits to a state with a different total angular moment $J = 1/2$ having no Stark structure rather than to a state with a different projection of the same angular moment. Thus, the experimental study of such three-body resonances should be much simpler, since in this case two-body resonance is completely absent.

Our numerical calculations using the example of the three-body Förster resonance $3 \times 70P_{3/2}(|M|=1/2) \to 70S_{1/2} + 71S_{1/2} + 70P_{1/2}$ for three Rydberg Rb atoms in several spatial configurations showed that for not too strong interaction, when various interaction channels are well resolved in the spectra, it is possible to observe high-contrast population oscillations. Since such oscillations are accompanied by oscillations of the phase of the collective wave function of three atoms, it is possible to develop new schemes of three-qubit quantum gates controlled by an electric field based on them. This is of interest for quantum information processing with qubits based on neutral atoms in arrays of optical dipole traps.

*Acknowledgements.* This work was supported by the Russian Foundation for Basic Research (Grant No. 19-52-15010) and Novosibirsk State University.